\documentclass[aps,prl,twocolumn,superscriptaddress]{revtex4-1}
\usepackage{graphicx}
\usepackage{epstopdf}
\usepackage{amsmath}
\usepackage{amssymb}
\usepackage{natbib}
\usepackage[latin1]{inputenc}

\begin{document}
	
\title{Enhanced transport length of spin-helical Dirac fermions in disordered 3D topological insulators}

\author{J. Dufouleur}
\email[email: ]{j.dufouleur@ifw-dresden.de}
\affiliation{IFW Dresden, P.O. Box 270116, D-01171 Dresden, Germany}

\author{L. Veyrat}
\affiliation{IFW Dresden, P.O. Box 270116, D-01171 Dresden, Germany}

\author{B. Dassonneville}
\affiliation{IFW Dresden, P.O. Box 270116, D-01171 Dresden, Germany}

\author{C. Nowka}
\affiliation{IFW Dresden, P.O. Box 270116, D-01171 Dresden, Germany}

\author{S. Hampel}
\affiliation{IFW Dresden, P.O. Box 270116, D-01171 Dresden, Germany}

\author{P. Leksin}
\affiliation{IFW Dresden, P.O. Box 270116, D-01171 Dresden, Germany}

\author{B. Eichler}
\affiliation{IFW Dresden, P.O. Box 270116, D-01171 Dresden, Germany}

\author{O. G. Schmidt}
\affiliation{IFW Dresden, P.O. Box 270116, D-01171 Dresden, Germany}

\author{B. Büchner}
\affiliation{IFW Dresden, P.O. Box 270116, D-01171 Dresden, Germany}

\author{R. Giraud}
\email[email: ]{r.giraud@ifw-dresden.de}
\affiliation{IFW Dresden, P.O. Box 270116, D-01171 Dresden, Germany}
\affiliation{CNRS, Laboratoire de Photonique et de Nanostructures, route de Nozay, 91460 Marcoussis, France}

\date{\today}

\begin{abstract}
The transport length $l_\textrm{tr}$ and the mean free path $l_\textrm{e}$ are determined for bulk and surface states in a Bi$_2$Se$_3$ nanoribbon by quantum transport and transconductance measurements. We show that the anisotropic scattering of spin-helical Dirac fermions results in a strong enhancement of $l_\textrm{tr}$, which confirms theoretical predictions \cite{Culcer2010}. Despite strong disorder ($l_\textrm{e}\approx30$~nm), the long-range nature of the scattering potential gives a large ratio $l_\textrm{tr}/l_\textrm{e}\approx8$, likely limited by bulk/surface coupling. This suggests that the spin-flip length $l_\textrm{sf} \approx l_\textrm{tr}$ could reach the micron size in materials with a reduced bulk doping, even if due to charge compensation.    
\end{abstract}

\maketitle

Disorder gives an important limitation to realize spintronic devices based on 3D topological insulators (3DTIs). Due to spin-momentum locking, the spin memory of surface Dirac fermions is lost on the transport length scale $l_\textrm{tr}$ \cite{Pesin2012}. Yet, the spin-flip length $l_\textrm{sf} \approx l_\textrm{tr}$ can affect the spin signals induced at the interface between a ferromagnetic contact and a 3DTI, and it even becomes a critical parameter to consider in a spin-valve device. Despite evidences for spin-charge conversion or spin pumping effects in 3DTIs nanostructures \cite{Mellnik2014, *Fan2014, *Deorani2014, *Liu2015, Shiomi2014}, a quantitative analysis in terms of surface transport is difficult, due to bulk states contribution to spin signals or to Hall voltages induced by local stray fields generated by a ferromagnetic contact \cite{Vries2015}. Even for bulk-insulating samples, a strong reduction of the spin-charge conversion efficiency was found \cite{Shiomi2014}, showing that the intrinsic spin physics of topological surface states can hardly be understood based on phenomenological models. This calls for a quantitative evalutation of microscopic parameters in a disordered 3DTI, such as the elastic mean free path $l_\textrm{e}$ or the spin-flip length, determined by $l_\textrm{tr}$, inferred for both surface and bulk carriers from transport measurements. 

Contrary to the case of 2D topological insulators \cite{Koenig2007}, the backscattering of topological surface states (TSS) is restored in a disordered 3DTI via $N$-multiple scattering events, on a finite lengthscale given by $l_\textrm{tr}$. The ratio $l_{\textrm{tr}}/l_\textrm{e} \approx N$ provides a robust way to measure the anisotropy of the scattering process \cite{Akkermans2007}. For metallic bulk carriers or Dirac fermions in graphene, elecronic screening is very efficient so that the scattering properties are determined by short-range disorder, which results in a small ratio $(l_{\textrm{tr}}/l_\textrm{e} \lesssim 2)$ \cite{DasSarma1985, Hwang2008, *Monteverde2010}. For TSS however, the spin texture of their band structure induces a strongly anisotropic scattering. According to theory, this leads to a very large value of $l_{\textrm{tr}}/l_\textrm{e}$  that can exceed 20 \cite{Culcer2010}. So far, such a high ratio was only found in modulation-doped semiconductor heterostructures for which a high-mobility 2D electron gas interacts with long-range disorder \cite{DasSarma1985, Paalanen1983, *Coleridge1989, *Mancoff1996}. Although the anisotropic scattering of TSS in a disordered 3DTI was investigated by scanning tunneling experiments \cite{Roushan2009,  *Lee2009a, *Zhang2009a, *Alpichshev2010, *Okada2011, *Kim2014, *Yee2015}, no transport study ever reported the measure and comparison of $l_\textrm{tr}$ and $l_\textrm{e}$.

In this letter, we report on such measurements and study the ratio $l_\textrm{tr}/l_{e}$ for all charge carriers contributing to the conductance of a single-crystalline Bi$_2$Se$_3$ nanoribbon with a density of Se vacancies in the $10^{19}$~cm$^{-3}$ range. Values of the carrier density $n$ and $l_\textrm{e}$ are extracted from Shubnikov-de Haas oscillations (SdHO) measurements, while $l_\textrm{tr}$ is inferred from transconductance measurements and, independently, from quantum transport experiments. Whereas similar values of $l_\textrm{e}$ are found for both bulk and surface states (same disorder), the values found for $l_\textrm{tr}$ differ by about one order of magnitude. The strong enhancement of the transport length of surface Dirac fermions, with a large ratio $l_\textrm{tr}/l_\textrm{e}\approx8$, confirms the long-range nature of the scattering potential, as expected from theory for spin-helical Dirac fermions \cite{Culcer2010}. Importantly, the resulting value obtained for the spin-flip length $l_\textrm{sf}^{\textrm{SS}} \approx l_\textrm{tr}^{\textrm{SS}} \approx 200$ nm shows that spin signals in a lateral spin valve based on Bi$_2$Se$_3$ can only be observed in the short-channel limit. Besides, we show that the long phase coherence length of topological surface states $l_\varphi^{\textrm{SS}}$, in agreement with previous reports in the diffusive regime \cite{Wang2011, *Kim2011b, *Li2015, *Ockelmann2015}, also results from the enhancement of $l_\textrm{tr}^{\textrm{SS}}$.

\begin{figure}[!h]
    \centering
    \includegraphics[width=\columnwidth]{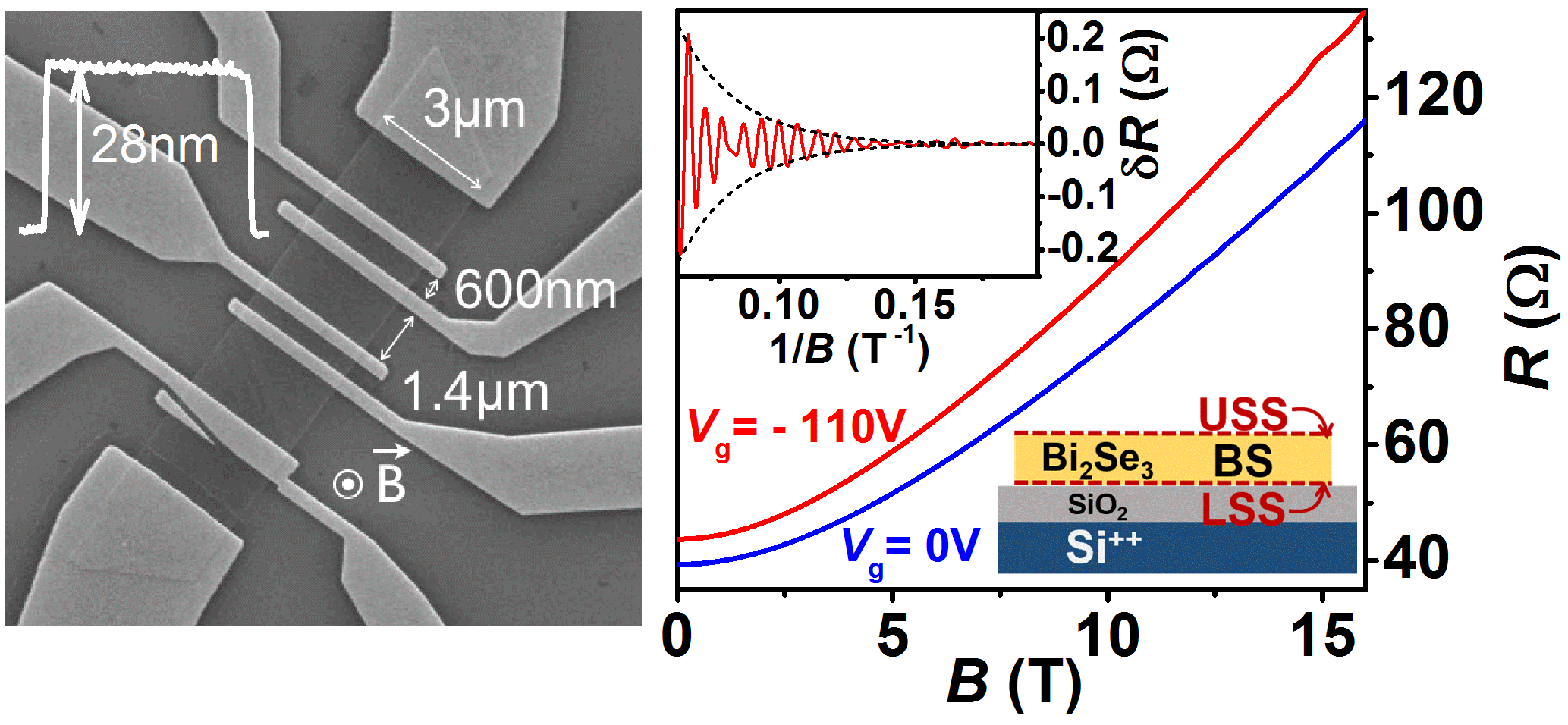}
    \caption[width=\columnwidth]{Left: a scanning electron microscope picture of the Bi$_2$Se$_3$ nanoribbon with a atomic force microscopy trace that indicates a flat surface with a thickness $t = 28$ nm. Right: magnetoresistance measured at $T=4$K for the long distance ($L=1.4 \mu$m) at $V_\textrm{g}=0$ V (blue) and $V_\textrm{g}=-110$ V (red). Inset: SdHO are evidenced after background substraction ($V_\textrm{g}=-110$ V).}
    \label{fig1}
\end{figure}

Single-crystalline Bi$_2$Se$_3$ nanostructures were grown by vapor transport directly on a Si$^{++}$-substrate covered with 300 nm of thermally grown SiO$_2$ \cite{Nowka2015, Veyrat2015}. Ohmic Cr(10 nm)/Au(100 nm) contacts were prepared on a nanoribbon ($L=20$ $\mu$m, $W=3$ $\mu$m, $t=28$ nm), by e-beam lithography and a metal lift-off technique, doing a wet etching in diluted HCl prior to the evaporation. The distances between contacts for the two measured pairs are 600 nm and 1.4 $\mu$m (Fig \ref{fig1}). A variable-temperature insert was used for measurements down to 2~K in a perpendicular magnetic field up to 16 T. Very low temperature measurements were performed in a Kelvinox 300 Oxford $^3$He/$^4$He dilution fridge with an electronic base temperature of about 30~mK, fitted into the bore of a 3D 2T-vector magnet with a 6 T perpendicular field. Transport properties were measured by standard lock-in techniques, using low-enough voltages so as to avoid electron heating.

\begin{figure}[!h]
	\centering
	\includegraphics[width=\columnwidth]{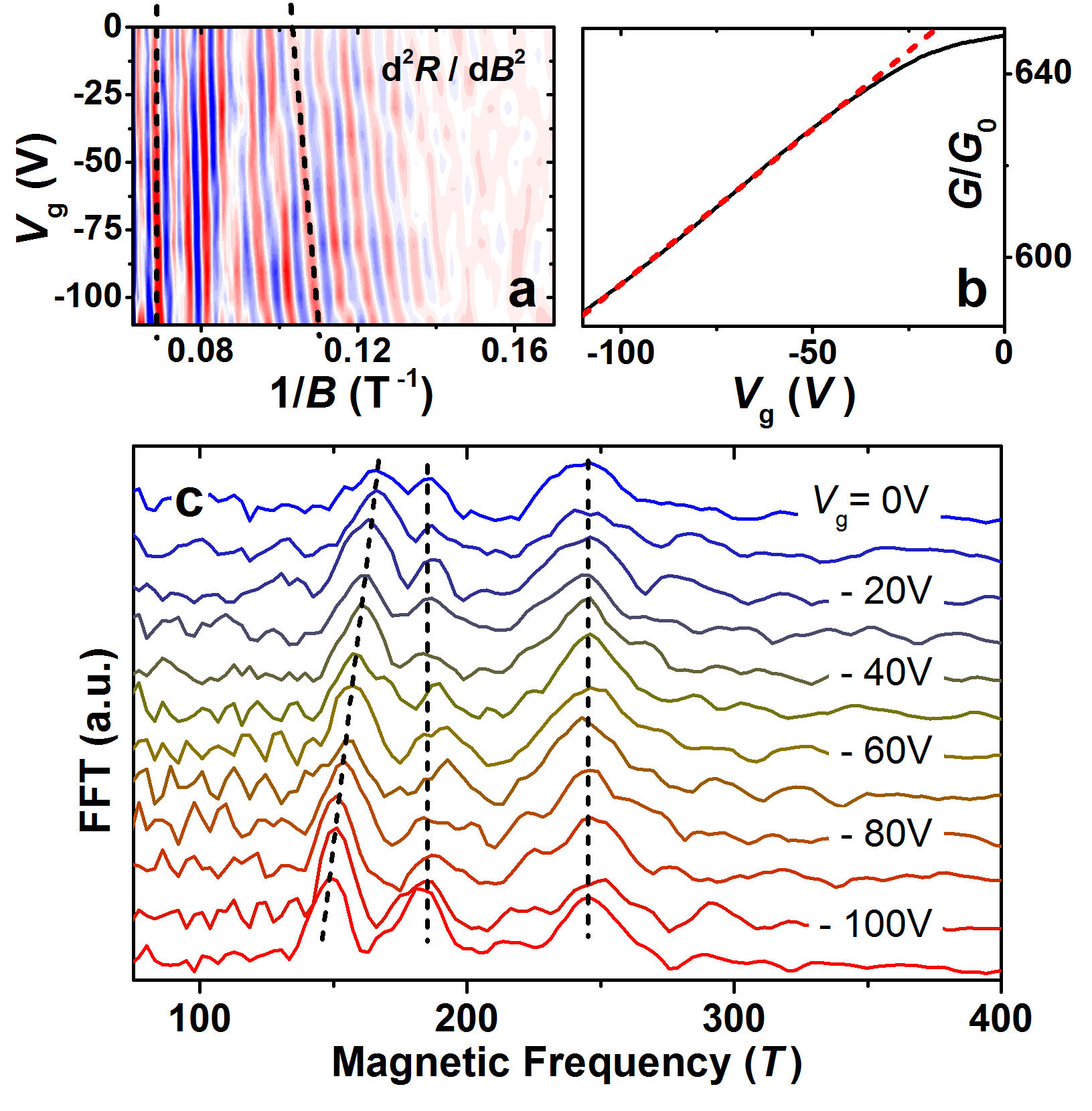}
	\caption[width=\columnwidth]{a) $d^2R/dB^2(1/B)$ for the long distance contact at $T=4$K and for different gate voltages ranging from 0 V to -110 V. Dashed lines indicate gate dependent and gate independent oscillations. b) Transconductance of the same contact with a theoretical fit derived from equations \ref{Conductance} and \ref{Conductivity} (red dashed line). c) FFT of the SdHO revealing three magnetic frequencies and their dependence with the gate voltage.}
	\label{fig2}
\end{figure}

We first focus on the measurement at high temperature (4.2 K $\leq T \leq $300 K). The sample shows a metallic behavior ($dR/dT > 0$) with $R_{\textrm{300K}}/R_{\textrm{4K}} \approx 2.2$ \cite{Veyrat2015}. At $T=4.2$ K, a strong magneto-resistance is observed when sweeping the perpendicular magnetic field up to 16T (Fig.\ref{fig1}), indicating a rather low electronic density, and tiny SdHO can be observed ($dR/R \sim 0.25\%$). These can be best studied from the second derivative of the magnetoresistance. A plot as a function of $1/B$ reveals two different families of periodic oscillations (Fig.\ref{fig2}a). The first one is visible at high magnetic fields ($B \gtrsim 11$ T) and has a short period $\Delta(1/B)\sim 0.005$ T$^{-1}$. The second one is visible already at lower magnetic fields (for $B \gtrsim 6.5$ T) and its period is slightly larger: $\Delta(1/B)\sim 0.007$ T$^{-1}$. A fast Fourier transform (FFT) of $d^2R/dB^2$ evidences a third magnetic frequency very close to the low-frequency one (Fig.\ref{fig2}c). The three different frequencies correspond to the three different charge populations contributing to the conductance (upper surface states - USS, lower surface states - LSS and bulk states - BS, respectively labelled "b", "u" and "l" below).

In order to identify each charge population, we studied their response to an electric field induced by a back-gate voltage. 
Only the lowest magnetic frequency peak is gate sensitive (Fig.\ref{fig2}c), so that it can be assigned to the LSS \cite{Sacepe2011}. From the temperature dependence of the low-field oscillations ($B<12$ T), which contribute only to the two lowest magnetic frequency peaks in the FFT, we found an effective mass $m^* \approx 0.15 \times m_\textrm{e}$ that could correspond to either bulk quasi-particles ($m^* \approx 0.124 \times m_\textrm{e}$ \cite{Koehler1973}) or TSS quasi-particles with an electronic density $n=4.1 \times 10^{12}$ cm$^{-2}$  ($m^*=2\hbar\sqrt{\pi n}/v_\textrm{F} =0.15 \times m_\textrm{e}$). For the high-field oscillations ($B>12$ T) however, $m^* \approx 0.2 \times m_\textrm{e}$ is significantly larger than the bulk effective mass but agrees very well with the value calculated for TSS with the corresponding electronic density $n=5.9 \times 10^{12}$ cm$^{-2}$ ($m^*=0.2 \times m_\textrm{e}$). We thus assign the high magnetic frequency peak to the USS and the last peak to BS \cite{Veyrat2015}.

Consequently, the different charge densities are $n_\textrm{b} = 1.4 \times 10^{19} $ cm$^{-3}$, $n_\textrm{l} = 4.1 \times 10^{12} $ cm$^{-2}$ and $n_\textrm{u} = 5.9 \times 10^{12} $ cm$^{-2}$. When a back-gate voltage $V_\textrm{g}$ is applied, $n_\textrm{l}$ changes according to the shift of the corresponding FFT peak: $dn_{\textrm{l}}/dV_\textrm{g} = 4.2 \times 10^9 $ cm$^{-2} $V$^{-1}$ (Fig.\ref{fig2}c). On the one hand, the shift of the LSS charge carrier density is about 10 times lower than the one induced by the capacitive effect so that the electric field is only partially screened by the LSS. On the other hand, the USS do not feel any electric field since their FFT peak is gate independent. Therefore, the electric field is fully screened by bulk carriers, over a distance given by the bulk Thomas-Fermi screening length $\lambda_{\textrm{TF}}=1/\sqrt{4 \pi e^2 / \epsilon_0 \epsilon_\textrm{r} \times \partial n / \partial \mu} \approx 2$ nm, where $\epsilon_0 \epsilon_\textrm{r}$ is the permittivity and $\partial n / \partial \mu$ the density of states. The thickness $t$ of the nanoribbon being much larger than $\lambda_{\textrm{TF}}$, the Fermi energy remains unaffected by a back gate voltage except close to the LSS so that other FFT peaks do not depend on $V_\textrm{g}$. The charge modification induced into the bulk by the capacitive effect is given by $dn_{\textrm{b}}/dV_\textrm{g}=(C/e-dn_{\textrm{l}}/dV_\textrm{g})/t=2.5 \times 10^{16} $ cm$^{-3} $V$^{-1}$ with $C$ the surface capacitance of the silicon oxide.

Based on the FFT data shown in Fig. \ref{fig2}, it is possible to extract the mean free path of all carriers independently by applying a band-pass filter in order to separate the SdHO associated to the three different magnetic frequencies at $V_\text{g}=-110$ V for which they are best separated. A Dingle plot for each charge carrier population gives the quantum lifetime and the corresponding mean free path for each type of carrier (Fig.\ref{fig1}). We find $l_\textrm{e}$=21 nm for BS and 28 nm for LSS and USS. Those values are rather close to each other, which points to a similar disorder, as expected for an homogeneous disorder related to Se vacancies. This value of $l_\textrm{e}$ is reasonable since the distance between two Se vacancies that act as donors in Bi$_2$Se$_3$ is about 5 nm.

The transport lengths $l_\textrm{tr}$ are extracted from transconductance measurements (Fig.\ref{fig2}b). The conductance is the sum of the bulk and surface states contributions:
\begin{equation}
    \label{Conductance}
    G=(\sigma_\textrm{b} \times t+\sigma_{\textrm{l}}+\sigma_{\textrm{u}})\times W/L
\end{equation}
where $\sigma_\textrm{b}$, $\sigma_{\textrm{l}}$, $\sigma_{\textrm{u}}$ are the different conductivities and $t$, $W$ and $L$ are respectively the thickness, the width and the length of the nanoribbon between the contacts we considered. Each conductivity is related to a diffusion coefficient $D$ and hence to a $l_\textrm{tr}$ through the relation:
\begin{equation}
    \label{Conductivity}
    \sigma=e^2 (\partial n/\partial \mu) D=e^2 (\partial n/\partial
    \mu) v_\textrm{F} l_\textrm{tr}/d
\end{equation}
where $d$ is the transport dimensionality. Since for the bulk $d_b=3$ $(\Leftrightarrow t \gtrsim l_\textrm{tr})$, it is more convenient to use the equivalent Drude formula: $\sigma=n e^2 l_\textrm{tr} / v_\textrm{F} m^*$. For TSS, we have $\partial n/\partial \mu= (\hbar v_\textrm{F})^{-1}\sqrt{n/\pi}$ with $v_\textrm{F}$ the Fermi velocity ($v_\textrm{F} = 5 \times 10^5$ m.s$^{-1}$ \cite{Xia2009}). In this case, $l_{\textrm{tr}}$ depends on the charge density and we have $l_{\textrm{tr}}^\textrm{l,u}=l_0\sqrt{n_\textrm{l,u}/n_0}$, where $l_0$ is the transport length at $n_{\textrm{l,u}}=n_0=n_{\textrm{l}}(V_\textrm{g}=0$ V$)$ \cite{Culcer2010}. Assuming the same disorder for LSS and USS, $l_{\textrm{tr}}$ can be expressed in the same way for both surface states and differences in $l_{\textrm{tr}}$ only come from the different charge densities.
 
Only two parameters are unknown: the transport length of the bulk $l_{\textrm{tr}}^\textrm{b}$ and the one of the TSS $l_0$. Using equations (\ref{Conductance}) and (\ref{Conductivity}) and taking into account the gate dependence of the charge densities, we can fit the transconductance with only $l_{\textrm{tr}}^\textrm{b}$ and $l_0$ as free parameters. We focus our analysis between $V_g=-110V$ and $V_g=-40V$ where the transconductance is linear and we find $l_b=28$nm and $l_0=196$nm. This very long $l_{\textrm{tr}}^{\textrm{l,u}}$ is in agreement with the observation of a quasiballistic regime in a Bi$_2$Se$_3$ nanowire of perimeter $L_p = 280$nm \cite{Dufouleur2013}. Such values explain the large contribution of the surface states to the total conductance of the nanoribbon, since $G_{\textrm{l}}+G_{\textrm{u}} \approx 1.2 \times G_{\textrm{b}}$. On the contrary, the contribution of the surface states to the transconductance is relatively small: $dG_\textrm{l}/dV_\textrm{g} \approx 0.2 \times dG_\textrm{b}/dV_\textrm{g}$.

Remarkably, even for a similar degree of disorder indicated by close values of $l_\textrm{e}$ for TSS and BS, the transport length are very different as indicated by the ratios $l_\textrm{tr}/l_\textrm{e}$ of 1.3, 7.0 and 8.4 for BS, LSS and USS. As expected from theory, this ratio is close to unity for massive spin-degenerate fermions in the bulk \cite{DasSarma1985}. It is much larger for TSS, due to anisotropic scattering, as predicted for spin-helical Dirac fermions in the presence of screened charged impurities \cite{Culcer2010}. Still, the measured ratio is significantly smaller than the theoretical prediction ($\approx 20$). This is probably due to a finite coupling between TSS and BS, so that we expect both the transport length and the spin-flip length to reach the micron-size limit if the density of impurities could be reduced to $10^{17}$~cm$^{-3}$. 

We now present quantum interference experiments that allow us to infer the phase coherence length $l_\varphi^\textrm{b}$ of bulk charge carriers. Importantly, those measurements provide a way to access the value of $l_{\textrm{tr}}^\textrm{b}$ independent from the previous approach. At very low temperature, quantum interferences between many different electronic trajectories induce quantum corrections to the classical Drude conductance. The interference pattern can be changed by tuning either the magnetic field or the Fermi energy thanks to the gate voltage. The amplitude of such reproducible universal conductance fluctuations (UCF) of the conductance $\delta G$ should approach the quantum of conductance $G_0=e^2/h$ for a fully coherent sample ($L \lesssim l_\varphi$). For $l_\varphi \lesssim L$, averaging over different disorder configurations reduces the amplitude of the UCF \cite{Akkermans2007}.

For a given Fermi energy $E_\textrm{F}(V_\textrm{g})=E$, we calculate the correlation function of the UCF along the $B$-axis 
$F_{\textrm{E}}(B)=\int \delta G(B')\delta G(B'+B)dB'$. We extract the correlation field $B_\textrm{c}$ corresponding to the half width at half maximum from the averaged correlation function $\langle F_{\textrm{E}}(B)/F_\textrm{E}(0)\rangle_\textrm{E}=1/\Delta E \int F_\textrm{E}(B)/F_\textrm{E}(0)dE$. It can be shown that $B_c \sim \phi_0 / l_\varphi^2$ where $\phi_0=h/e$ is the quantum of flux and $h$ is the Planck constant\cite{Lee1987}. Providing that the gate voltage dependence of the Fermi energy is known, it is possible to extract in a similar way a correlation energy $E_\textrm{c}$ related to $l_\varphi$ and to the diffusion coefficient $D=v_\textrm{F} l_\textrm{tr}/d$ if $l_\varphi <L$ \cite{Lee1985a,Lee1987}:

\begin{equation}\label{CorrelationEnergy}
    E_\textrm{c}=(\pi / 2) hD / {l_\varphi}^2
\end{equation}

The relation between the Fermi energy and the carrier density of the lower surface state $E_\textrm{F}^\textrm{l}$ and $n_\textrm{l}(V_\textrm{g})$ is given by $E_\textrm{F}^\textrm{l}=hv_\textrm{F}\sqrt{n_\textrm{l}/\pi}$. The gate dependence of $E_\textrm{F}^\textrm{l}$ gives the gate dependence of the bulk Fermi energy $E_\textrm{F}^\textrm{b} (V_\textrm{g})$ at the interface with the lowest surface states. The electric field is screened by bulk charge carriers for distances from the lower interface longer than $\lambda_\textrm{TF}$ and $E_\textrm{F}^\textrm{b}$ is not gate dependent anymore. We consider here that $B_\textrm{c}$ and $E_\textrm{c}$ are dominated by bulk quasi-particles trajectories. Hence, the number of quasi-particle's trajectories is about one order of magnitude larger into the bulk than into the TSS. Bulk thus governs UCF properties providing that: 1) TSS and BS are coupled to each other \cite{Steinberg2011, Li2015}, and 2) the phase coherence length of the bulk $l_\varphi^\textrm{b}$ is not negligible as compared to that of the surface states $l_\varphi^\textrm{l,u}$.

\begin{figure}[!h]
	\centering
	\includegraphics[width=\columnwidth]{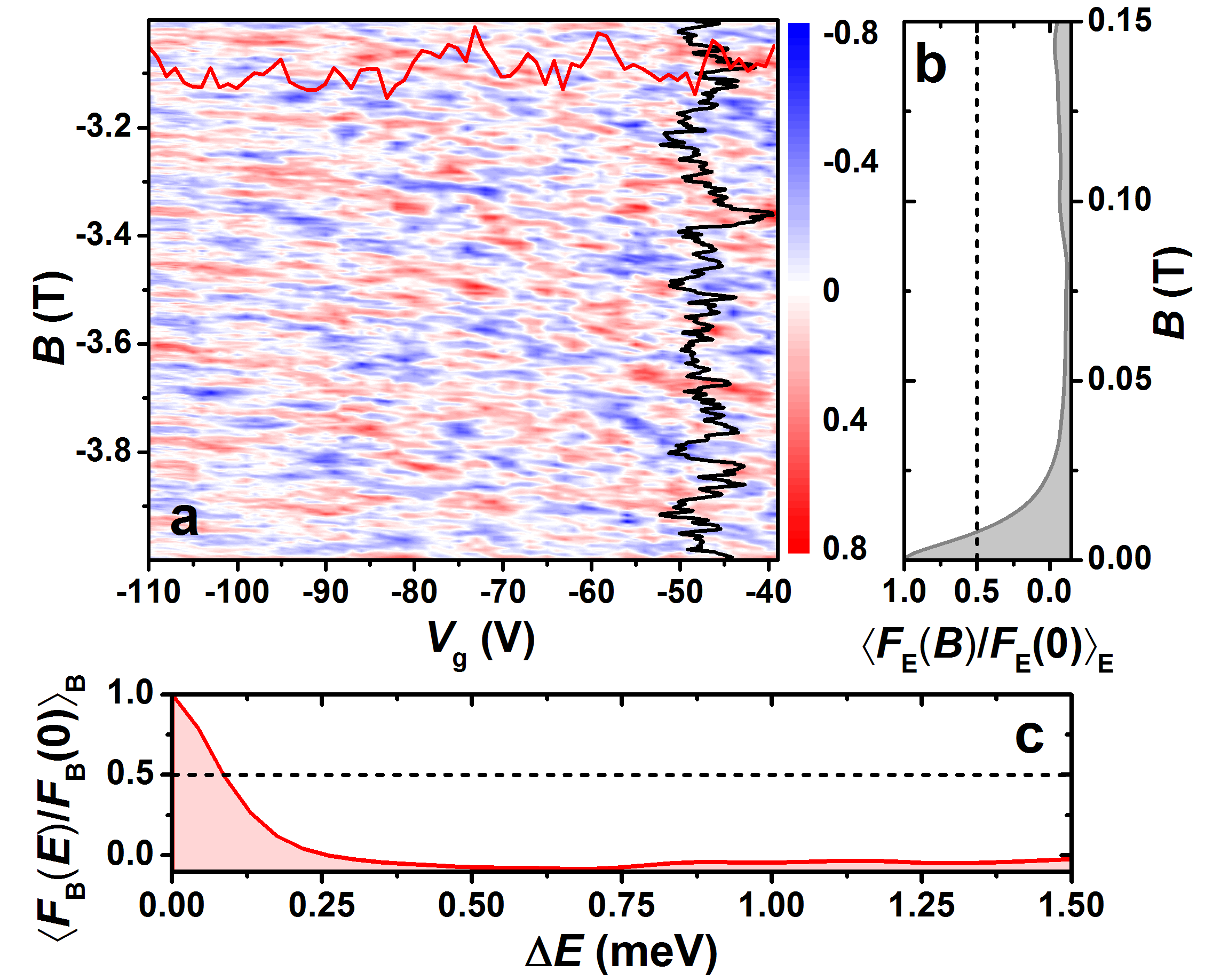}
	\caption[width=\columnwidth]{a) Conductance fluctuations $\delta G (V_\textrm{g},B)/G_0$ measured at $T=50$ mK in the nanoribbon for a distance between the contacts $L = 1.4 \mu$m (see Fig.\ref{fig1}). A smooth background is removed in order to get rid of classical transconductance effect. The red trace corresponds to $\delta G (V_\textrm{g})$ at $B=-3$ T and the black one to $\delta G (B)$ at $V_\textrm{g}=-39$ V. b) Field dependence of the UCF correlation function averaged over the full energy range measured and c) the energy correlation function averaged over the full magnetic field range.}
	\label{fig3}
\end{figure}

At $T=50$ mK, we extract a correlation field of about 8 mT which corresponds to a $l_\varphi^\textrm{b} \approx 720$ nm for a distance between the contacts $L=1.4$ $\mu$m. The other contact configuration ($L=600$ nm) leads to $l_\varphi^b \approx 520$ nm so that we take an average value of the phase coherence length $l_\varphi^\textrm{b}=620$ nm. The typical correlation energy is $E_\textrm{c} \approx 85$ $\mu$eV. The same value is found for the two pairs of contacts. From $E_\textrm{c}$ and $l_\varphi^\textrm{b}$, we derive a bulk transport length $l_\textrm{tr} \approx 31$ nm, in good agreement with the value inferred from transconductance measurements.

Finally, taking the value of $l_\varphi^\textrm{b} \approx 620$ nm, it is possible to estimate a lower bound for $l_\varphi^\textrm{SS}$. For such a wide ribbon, we can assume that the phase coherence time $\tau_\varphi = {l_\varphi}^2 / D$ is the same for BS and TSS and that the only difference in $l_\varphi$ comes from the different diffusion coefficients for BS ($D_\textrm{b}$) and TSS ($D_\textrm{SS}$). We find $l_\varphi^{\textrm{SS}}=\sqrt{D_{\textrm{SS}}/D_\textrm{b}} \times l_\varphi^\textrm{b}$ leading to $l_\varphi^{\textrm{SS}} \approx 1.3$ $\mu$m. This ratio $l_\varphi^{\textrm{SS}} / l_\varphi^\textrm{b}$ is very similar to the one recently inferred in Bi$_2$Te$_2$Se nanostructures from weak localization measurements \cite{Li2015}. The long phase coherence length of TSS results here from the enhancement of $l_\textrm{tr}$. We note that the value of $l_\varphi^{\textrm{SS}}$ can be further increased in reduced dimensionality due to the longer $\tau_\varphi$ for quasi-ballistic 3D TI nanostructures, as reported for Bi$_2$Se$_3$ quantum nanowires \cite{Dufouleur2013}.

Our results show a strong enhancement of the transport length for topological surface states in a disordered 3D topological insulator, contrary to the case of massive quasi-particles or of spin-degenerate Dirac fermions in graphene \cite{Monteverde2010}. This is a direct signature of the anisotropic scattering of spin-helical Dirac fermions. The large ratio $l_\textrm{tr}/l_\textrm{e} \approx 8$ reveals that the disorder is dominated by screened charged impurities instead of a short-range disorder \cite{Culcer2010}. The large transport length also explains the enhancement of the phase coherence length of the surface states. Importantly, the long $l_{\textrm{tr}}^{\textrm{SS}}$ and $l_\varphi^{\textrm{SS}}$ are key parameters to consider for the development of new ballistic quantum devices and spintronic devices made out of disordered 3D-topological insulator nanostructures. In highly-disordered Bi$_2$Se$_3$ nanostructures, our quantitative estimate of a rather long transport length $l_\textrm{tr}^{\textrm{SS}} \approx 200$ nm confirms that the spin-flip length could reach the micron size in disordered 3D topological insulator nanostructures with a weaker disorder \cite{Koirala2015} or a reduced bulk doping (a defect density of about $10^{17}$~cm$^{-2}$ would increase all lengths by a factor five), even if due to charge compensation (decoupling between surface and bulk states).

\subsection*{Acknowledgments}
J.D. gratefully acknowledges the support of the German Research Foundation DFG through the SPP 1666 Topological Insulators program and B.D. thanks the Alexander von Humboldt foundation for its financial support.


\bibliographystyle{apsrev4-1}
\bibliography{AnisotropicScattering}



\end{document}